\documentstyle[aps,prl,twocolumn,floats,epsfig]{revtex}
\begin{document}

\newcommand{\tbox}[1]{\mbox{\tiny #1}}
\newcommand{\half}{\mbox{\small $\frac{1}{2}$}}
\newcommand{\sfrac}[1]{\mbox{\small $\frac{1}{#1}$}}
\newcommand{\mbf}[1]{{\mathbf #1}}


\title{Failure of random matrix theory 
to correctly describe quantum dynamics} 

\author
{Tsampikos Kottos$^1$ and Doron Cohen$^2$ \\
\footnotesize
$^1$
Max-Planck-Institut f\"ur Str\"omungsforschung,
37073 G\"ottingen, Germany \\
$^2$
Department of Physics, Harvard University, Cambridge, MA 02138 \\
}

\date{April 2001}

\maketitle


\begin{abstract}
Consider a classically chaotic system which is described 
by a Hamiltonian ${\cal H}_0$. At $t=0$ the Hamiltonian 
undergoes a sudden-change ${\cal H}_0 \mapsto {\cal H}$. 
We consider the quantum-mechanical spreading of the evolving
energy distribution, and argue that it cannot be analyzed 
using a random-matrix theory (RMT) approach. RMT can be 
trusted only to the extend that it gives trivial results 
that are implied by first-order perturbation theory.  
Non-perturbative effects are sensitive to the underlying 
classical dynamics, and therefore the $\hbar\rightarrow 0$ 
behavior for effective RMT models is strikingly different 
from the correct semiclassical limit.    
\end{abstract}    
 
\ \\
  

Consider a system whose total Hamiltonian is ${\cal H}={\cal H}(Q,P;x)$, 
where $(Q,P)$ is a set of canonical coordinates, and $x$ is a 
constant parameter. We assume that the preparation and the representation 
of the system are determined by the Hamiltonian 
${\cal H}_0={\cal H}(Q,P;x_0)$, and that both ${\cal H}_0$ and ${\cal H}$ 
generate classically chaotic dynamics of similar nature. 
Moreover, we assume that $\delta x\equiv (x{-}x_0)$ 
is {\em classically small}, meaning that it is possible to apply 
linear analysis in order to describe how the energy 
surfaces ${\cal H}(Q,P;x)=E$ are deformed as a result of changing 
the value of $x$. Physically, going from ${\cal H}_0$ to 
${\cal H}$ may signify a change of an external field, or switching 
on a perturbation, or sudden-change of effective-interaction 
(as in molecular dynamics). Quantum mechanically, we can use a basis 
where ${\cal H}_0 = \mbf{E}_0$ has a diagonal representation, while 
\begin{eqnarray} \label{e1} 
{\cal H} \ \  = \ \ \mbf{E}_0 \ + \ \delta x \ \mbf{B}
\end{eqnarray}
For reasonably small $\hbar$, it follows from general semiclassical 
considerations \cite{mario}, that $\mbf{B}$ is a {\em banded matrix}.  
Generically, this matrix {\em looks random}, as if its off-diagonal 
elements were {\em independent} random numbers. 

It was the idea of Wigner \cite{wigner} forty years ago, 
to study a simplified model, where the Hamiltonian 
is given by Eq.~(\ref{e1}), and where $\mbf{B}$ 
is a Banded Random Matrix (BRM) \cite{felix}. 
This approach is attractive both analytically and numerically. 
Analytical calculations are greatly simplified by the 
assumption that the off-diagonal terms can be treated as 
independent random numbers. Also from numerical point 
of view it is quite a tough task to calculate the true 
matrix elements of the $\mbf{B}$ matrix. It requires 
a preliminary step where the chaotic ${\cal H}_0$ is 
diagonalized. Due to memory limitations one ends up with 
quite small matrices. We can think of Eq.~(\ref{e1}) as 
describing fictitious motion on a lattice. For the model 
below (Eq.~\ref{e2}) we were able to handle $N=5000$ sites 
maximum. This should be contrasted with BRM simulations, 
where using self-expanding algorithm \cite{slfex} we were able to 
handle $N=100000$ sites along with significantly reduced CPU time.

However, the applicability of the RMT approach is a matter 
of conjecture. Obviously this conjecture should be tested. 
To be more specific, one should be aware that there is a hierarchy 
of challenges where the applicability of the RMT conjecture should 
be tested. Namely: The study of spectral statistics; 
The study of eigenstates; The study of quantum dynamics. 
While the issue of spectral statistics has become a major 
subject in "quantum chaos" studies \cite{smpl}, the two other 
issues are barely treated. In a previous study \cite{lds} we have 
demonstrated that the RMT approach is capable of giving the 
right qualitative picture of the parametric evolution of the 
eigenstates. As $\delta x$ is increased the eigenstates 
of Eq.~(\ref{e1}) change in a qualitative agreement with Wigner's theory. 
Still, RMT fails to capture non-universal system-specific features.

In this Letter we turn to the study of quantum {\em dynamics}. 
Here we are going to end up with a much more alarming claim. 
Namely, the RMT approach fails to give the correct dynamical picture. 
RMT can be trusted only to the extend that it gives trivial results 
that are implied by first-order perturbation theory.  
Non-perturbative effects are sensitive to the underlying 
classical dynamics, and therefore the $\hbar\rightarrow 0$ 
behavior for effective RMT models is strikingly different 
from the correct semiclassical limit. 
In this Letter we are going to establish the failure of RMT for 
the case of dynamics which (for $t>0$) is generated by a time 
independent Hamiltonian. This we hope paves the way towards making 
an analogous statement regarding the response of driven systems \cite{rsp}.

\begin{figure}[t] 
\centerline{\epsfig{figure=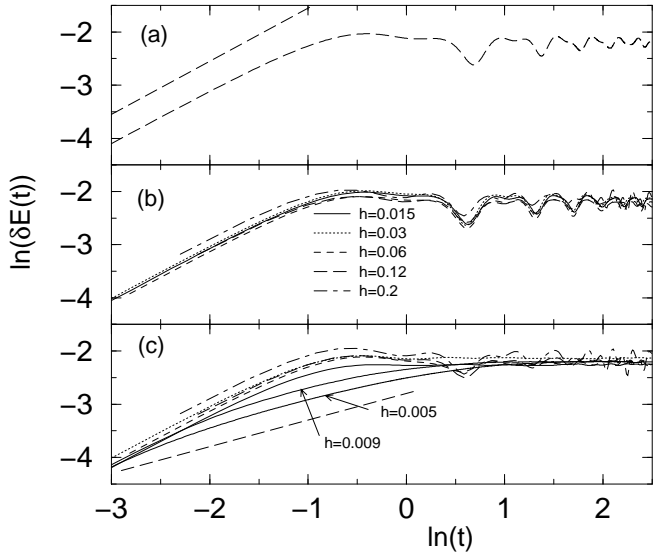,width=\hsize}}
\vspace{.1in}
\caption{
(a) The classical energy spreading as a function of time. 
(b) The QM spreading for the 2DW Hamiltonian; 
(c) The QM spreading for the EBRM Hamiltonian. 
The energy in these simulations is $E\sim 3$, 
and $\delta x = 0.2123$.  
In (a) we see a crossover from ballistic spreading 
($\delta E  \propto t$) to saturation ($\delta E \sim \mbox{const}$).
Only one time scale ($\tau_{\tbox{cl}} \sim 1$) is involved. 
The light dashed line has slope $1$ and is drawn to guide the eye. 
In (b) we see that the classical behavior is approached 
as $\hbar\rightarrow 0$. In (c) we see the opposite 
trend: as $\hbar\rightarrow 0$ an intermediate stage 
of diffusion ($\delta E  \propto \sqrt{t}$) develops.  
(Here the light dashed line has slope $1/2$).
Different lines correspond to different values 
of $\hbar$ as in (b), and additional curves 
($\hbar=0.009,0.005$) have been added.
}
\end{figure}

\begin{figure}[t] 
\centerline{\epsfig{figure=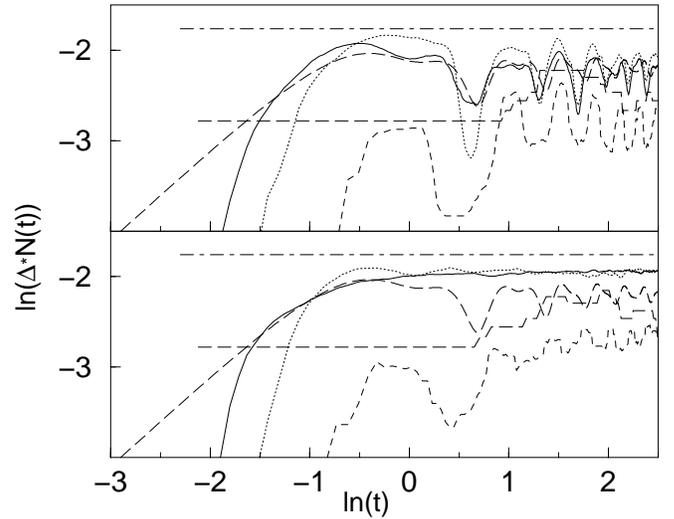,width=\hsize}}
\vspace{.1in}
\caption{
The width $\Delta \times N(t)$ as a function of time.  
Different lines correspond to different $\hbar$-values 
as in Fig.~1b. The heavy dashed line is the 
classical $\delta E(t)$. Having separation of scales 
($\Delta {\times} N(t) \ll \delta E(t)$) is an 
indication for having a {\em perturbative} spreading profile.  
The upper panel is for the 2DW Hamiltonian 
while the lower panel is for the EBRM model. 
}
\end{figure}

In order to test the RMT conjecture, 
we are going to use the same `direct' approach and the same model  
as in Ref.\cite{lds}. For the first time we are going to compare 
the {\em dynamics} which is generated by a `physical' Hamiltonian, 
with the corresponding dynamics that is obtained from an 
effective BRM model (EBRM). 
The latter is constructed by taking the matrix $\mbf{B}$ of 
the `physical' Hamiltonian, and then randomizing the signs of its 
off-diagonal elements. Such operation destroys any 
correlations between the matrix elements of $\mbf{B}$,  
while keeping the band profile unaffected \cite{ewbrm}.
We study the Hamiltonian \cite{lds}
\begin{eqnarray} \label{e2} 
{\cal H}(Q,P;x) = \half(P_1^2{+}P_2^2 + Q_1^2{+}Q_2^2)  
+ x \cdot Q_1^2 Q_2^2
\end{eqnarray}
with $x=x_0+\delta x$ and $x_0=1$. This Hamiltonian describes the 
motion of a particle in a 2D well (2DW). The units are chosen 
such that the mass is equal to one, the frequency for small 
oscillations is one, and for $\delta x=0$ the coefficient of the 
anharmonic term is also one. The energy $E$ is the only 
dimensionless parameter of the classical motion. Our numerical 
study is focused on an energy window around $E \sim 3$ where 
the motion is mainly chaotic. Upon quantization we have 
a second dimensionless parameter, which is the scaled $\hbar$.
Associated with $\hbar$ are two energy scales. 
One is the mean level spacing $\Delta \propto \hbar^d$ 
with $d=2$, and the other is the bandwidth $\Delta_b \propto \hbar$. 
The second scale is further discussed below.

It is useful to define a fluctuating quantity ${\cal F}(t) \equiv 
-(\partial {\cal H}/\partial x)$ which for the 2DW model equals 
${\cal F}(t) = -Q_1^2(t) Q_2^2(t)$. The auto-correlation function 
of ${\cal F}(t)$ is denoted by $C(\tau)$. The associated correlation 
time is denoted by $\tau_{\tbox{cl}}$. The power spectrum of the 
fluctuations $\tilde{C}(\omega)$ is the Fourier transform of 
$C(\tau)$. The band profile of the matrix $\mbf{B}$ satisfies the 
semiclassical relation $|\mbf{B}_{nm}|^2 \approx (\Delta /(2\pi\hbar)) 
\tilde{C}((E_n{-}E_m)/\hbar)$.
See Fig.2 of \cite{lds} for numerical demonstration. 
It is implied by this relation that the bandwidth  
is $\Delta_b = 2\pi\hbar/\tau_{\tbox{cl}}$. 
For the quantum mechanical simulation the exact matrix 
$\mbf{B}$ has been calculated numerically. Memory constraints 
limit the maximum size ($N$) of the matrix that we can get. 
The EBRM Hamiltonian is obtained by randomizing the 
signs of the off-diagonal elements. A second, more `loose'  
strategy, is to generate the EBRM $\mbf{B}$ from scratch 
using the semiclassical band-profile as an input. 
The advantage of the latter strategy is that it opens 
the way to EBRM-model simulations with smaller $\hbar$, 
where the required $N$ is much larger.  We have verified 
that the latter strategy gives numerical results that 
agree with the sign-randomization approach.

The initial preparation is assumed to be microcanonical. 
This means, in the classical case, an ergodic distribution 
of initial `points' on the energy surface ${\cal H}_0(Q,P) = E$ 
with $E \sim 3$.
In the quantum mechanical case we start each simulation 
with an initial eigenstate $m$ that has an energy $E_m \sim E$ 
where $2.75 < E < 3.2$. The time dependent evolution is 
determined by Schrodinger Equation. The probability distribution 
after time $t$ is $P_t(n|m)$. An average over 
initial state ($m$) is taken in order to get the average 
profile $P_t(n-m)$.  
We characterize the evolving distribution using three
different measures. The variance is $M(t)=\sum_r r^2 P_t(r)$, 
or in energy units it is $\delta E(t) = \Delta \times \sqrt{M(t)}$.   
The width $N(t)$ is defined as the $r$~region that 
contains $50\%$ of the probability. In case that we have 
a spreading profile that is characterized by a single 
energy scale, it is implied that $N(t)$ and $\sqrt{M(t)}$ 
would be the same (up to a numerical factor). 
The survival probability is $P(t)=P_t(r{=}0)$. 
The results of the simulations are presented in the Figs~{1-4}. 
The analysis of these results is discussed below.

\begin{figure}
\centerline{\epsfig{figure=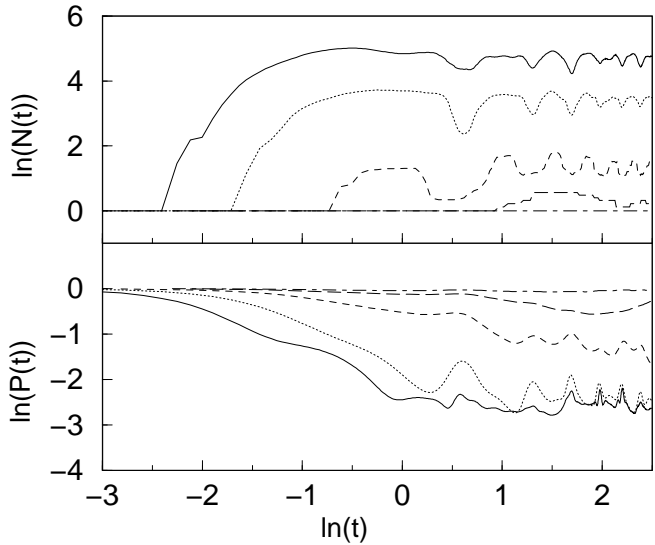,width=\hsize}}
\vspace{.1in}
\caption{
The width $N(t)$ and the survival probability $P(t)$ 
for the 2DW model.  Different lines correspond to 
different $\hbar$-values as in Fig.~1b. 
Having $N(t)=1$ or $P(t)\sim 1$ is the indication 
for having a {\em standard} perturbative spreading profile. 
}
\end{figure}

Taking ${\cal H}$ to be a generator for the classical 
dynamics, the energy $E(t) = {\cal H}_0(Q(t),P(t))$ fluctuates. 
The fluctuations are characterized by the correlation time
$\tau_{\tbox{cl}}$, and by an amplitude $\delta E_{\tbox{cl}}$.
The initial preparation is assumed to be a microcanonical distribution 
that is supported by the energy-surface ${\cal H}_0(Q,P)=E(0)$. 
For $t>0$, the phase-space distribution 
spreads away from the initial surface. `Points' of the evolving
distribution move upon the energy-surfaces of ${\cal H}(Q,P)$.
We are interested in the distribution of the energy $E(t)$
of the evolving `points'. 
It is easily argued that for short times 
this distribution evolves in a ballistic fashion. 
Then, for $t \gg \tau_{\tbox{cl}}$, 
due to ergodicity, a `steady-state distribution' appears, 
where the evolving `points' occupies an `energy shell' in phase-space. 
The thickness of this energy shell equals $\delta E_{\tbox{cl}}$.
Thus we have a crossover from ballistic energy spreading 
to saturation. The dynamics in the classical limit 
is fully characterized by the two classical 
parameters $\tau_{\tbox{cl}}$ and $\delta E_{\tbox{cl}}$.

A quantitative description of the classical spreading 
is easily obtained. A straightforward derivation leads to 
the following result for the spreading 
\begin{eqnarray} \label{e3} 
\delta E_{\tbox{cl}}(t) = \delta x \times \sqrt{2(C(0)-C(t))}
\end{eqnarray} 
As a particular result we get $\delta E_{\tbox{cl}} \equiv \delta E_{\tbox{cl}}(\infty)= 
\delta x \sqrt{2C(0)}$. The calculation of $\delta E_{\tbox{cl}}(t)$ for the model 
Hamiltonian is presented in Fig.~1a. 
It is implied by Eq.(\ref{e3}) that the spreading $\delta E(t)$, 
from semiclassical point of view, 
is just a property of the band-profile. Thus, one may get  
to the wrong conclusion that models with the same 
band-profile should lead to the same $\delta E(t)$, 
provided the off-diagonal elements look random.  
If this were the case, it would be implied that 
the EBRM model would be equivalent to the 2DW model as far as 
the spreading $\delta E(t)$ is concerned. Looking on Fig.1 
we see that this is not the case. As $\hbar\rightarrow 0$ 
the EBRM model further and further deviates from the 
(correct) semiclassical expectation.

In order to understand the observed results we would 
like to recall some of the theory of \cite{crs,lds}. 
We already said that upon quantization we have 
the two energy scales $\Delta \propto \hbar^d$ and 
and $\Delta_b \propto  \hbar$. Actually there is also 
a semiclassical energy scale $\Delta_{\tbox{SC}} \propto \hbar^{2/3}$. 
Associated with these energy scales are three parametric 
scales $\delta x_c^{\tbox{qm}} \ll \delta x_{\tbox{prt}} \ll \delta x_{\tbox{SC}}$, 
where the strong inequalities hold in the $\hbar\rightarrow 0$ limit. 
For the 2DW model, assuming $E\sim 3$ we have \cite{lds}
the estimates $\delta x_c^{\tbox{qm}} \approx 3.8*\hbar^{3/2}$ 
and $\delta x_{\tbox{prt}} \approx 5.3*\hbar$ 
and $\delta x_{\tbox{SC}} \approx 4*\hbar^{2/3}$. 
In the standard perturbative regime ($\delta x < \delta x_c^{\tbox{qm}}$)  
the eigenstates of Eq.(\ref{e1}) have a simple perturbative 
structure. In the extended perturbative regime 
($\delta x_c^{\tbox{qm}}  < \delta x < \delta x_{\tbox{prt}}$) 
the eigenstates of Eq.(\ref{e1}) have a core-tail 
structure that can be regarded as a generalization of 
Wigner's Lorentzian. In the non-perturbative regime 
($\delta x > \delta x_{\tbox{prt}}$) the eigenstates have a 
purely non-perturbative structure. Depending on whether 
it is the EBRM Hamiltonian or the 2DW Hamiltonian, this 
`ergodic' non-perturbative structure is either semicircle-like 
or semiclassical-like respectively. The semiclassical 
regime ($\delta x > \delta x_{\tbox{SC}}$) is contained in the 
non-perturbative regime. It is only there that we 
can trust detailed quantal-classical correspondence (QCC).

For the purpose of the present analysis it is convenient to specify 
the different regimes by regarding $\hbar$ as a free parameter. 
Thus, the standard perturbative regime is \mbox{$\hbar > C_{\tbox{cqm}}$}, 
the extended perturbative regime is \mbox{$C_{\tbox{prt}} < \hbar < C_{\tbox{cqm}}$}, 
and the non-perturbative regime is \mbox{$\hbar < C_{\tbox{prt}}$}. The latter 
contains the semiclassical regime \mbox{$\hbar < C_{\tbox{SC}}$}. Thus the 
semiclassical limit $\hbar\rightarrow 0$ is a non-perturbative limit.
In case of the 2DW model, the classical quantities 
are \mbox{$C_{\tbox{cqm}}=0.41*dx^{2/3}$} 
and \mbox{$C_{\tbox{prt}}=0.19*dx$} 
and \mbox{$C_{\tbox{SC}}=0.12*dx^{3/2}$}.
We have used in most of our numerical simulations 
$\delta x \sim 0.2$. Larger $\delta x$ may take us 
out of the classical linear regime. For this 
value of $\delta x$ we get $C_{\tbox{cqm}}=0.14$
and $C_{\tbox{prt}}=0.04$ and $C_{\tbox{SC}}=0.01$. 
The smallest $\hbar$ value that we could allow 
without having memory-overflow was $\hbar=0.015$.  
This means that we were able to access the 
non-perturbative regime, though the semiclassical 
regime was out of reach.

As explained in \cite{wbr} the essential features 
of the spreading behavior in the perturbative regimes 
can be analyzed using first order perturbation theory (FOPT). 
Since correlations between off-diagonal elements are 
not important for FOPT, it follows that the EBRM-model and 
the 2DW model should be trivially equivalent in such case. 
It is only in the non-perturbative regime where 
the question of their equivalence becomes non-trivial. 
In case of the standard BRM we have witnessed \cite{wbr}
in the non-perturbative regime a premature departure 
from ballistic behavior, and appearance of an intermediate 
diffusive stage. We observe essentially the same 
behavior in case of the EBRM Hamiltonian (Fig.~1c).   
But with the 2DW Hamiltonian (Fig.~1b) we do not have 
such an effect:  
As $\hbar \rightarrow 0$ the correspondence 
with the classical behavior becomes better and better. 
Thus our simulations demonstrate that having 
diffusion in the non-perturbative regime is an 
artifact of the RMT approach. 
In our previous work \cite{wbr} we did not have 
a numerical proof to support such a strong statement. 
There, all we were able to do, was to argue that RMT 
should fail in the deep semiclassical regime
($\hbar \ll C_{\tbox{SC}}$), 
thus leaving open the possibility for having an intermediate 
regime ($C_{\tbox{SC}} < \hbar < C_{\tbox{prt}}$) where RMT might 
be valid. As we see, our numerical results do not give 
any indication for the existence of such intermediate regime.
The failure of RMT happens {\em as soon as} we enter 
the non-perturbative regime ($\hbar < C_{\tbox{prt}}$).

\begin{figure}
\centerline{\epsfig{figure=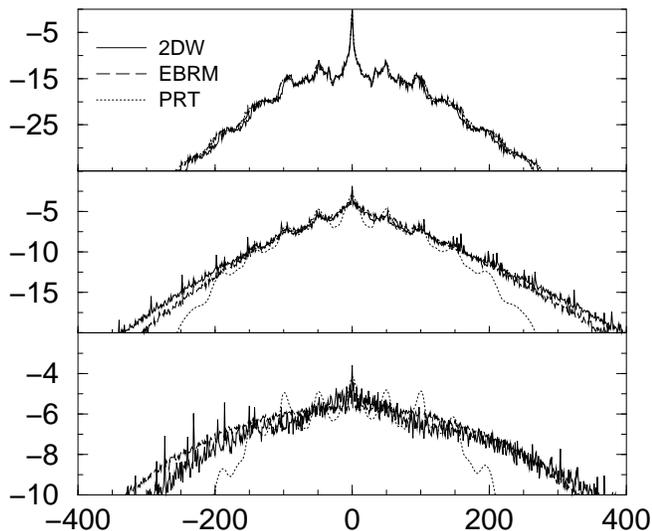,width=\hsize}}
\vspace{.1in}
\caption{
The spreading profile $P_{\infty}(r)$   
in representative cases. 
The upper panel is an example for 
a standard perturbative profile 
($P(t)\sim 1$). The middle panel 
is an example for a perturbative 
core-tail structure 
($\Delta {\times} N  \ll \delta E$). 
The lower panel is an example for 
an ergodic-like non-perturbative structure  
($\Delta {\times} N  \sim \delta E$).
}
\end{figure}

As we make $\hbar$ smaller there is no indication in 
Fig.~1 for entering a non-perturbative regime. 
In \cite{crs,lds,wbr} we have made the important 
distinction between {\em detailed} QCC and {\em restricted} QCC. 
The former pertains to the whole spreading profile, 
while the latter pertains only to the variance. 
Restricted QCC is a robust type of correspondence that 
does not require a `very small $\hbar$'. If we want 
to have an indication for the crossover to a non-perturbative 
behavior we should look on other measures, such as 
$N(t)$ of Fig.~2.
If the spreading were classical-like, it would imply 
that the spreading profile is characterized by a single 
energy scale. In such case we would expect 
that $N(t)$ and $\sqrt{M(t)}$ would be the same 
(up to a numerical factor). 
Indeed this is the case for the $\hbar<0.04$ runs of Fig.2. 
However, this is definitely not the case in the 
perturbative regimes, where we have a separation of energy 
scales $N(t) \ll \sqrt{M(t)}$. 
In the perturbative regimes $M(t)$ is determined 
by the tails, and it is insensitive to the 
size of the `core' region.  The width $N(t)$ 
constitutes a practical estimate for the latter (see Fig.3a).   
It is $N(t)=1$ for a {\em standard} perturbative profile, 
and  $1 \ll N(t) \ll \sqrt{M(t)}$  for a fully developed 
core-tail structure. An alternate way to identify 
a standard perturbative profile is via the survival 
probability $P(t)$. Indeed for $\hbar<0.14$ we 
see in Fig.3b that we have $P(t)\sim 1$.

The difference between the perturbative and the 
non-perturbative spreading profiles is further 
illustrated in Fig.4. Here we have plotted representative 
average saturation profiles, along with a comparison with 
the perturbative core-tail calculation (PRT for brevity). 
The saturation profile is given by the expression
$P_{\tbox{$\infty$}}(n|m)=
\sum_{n'} |\langle n(x_0)|n'(x)\rangle|^2 |\langle n'(x)|m(x_0)\rangle|^2$.  
It can be regarded as the auto-convolution of 
$P_{\tbox{E}}(n|m)= |\langle n(x)|m(x_0)\rangle|^2$. 
Thus the {\em average} saturation profile $P_{\tbox{$\infty$}}(r)$ 
is approximately related to the {\em average}  
local density of states $P_{\tbox{E}}(r)$.   
The latter has been analyzed in \cite{lds}.  
In Fig.3 we have calculated the PRT of $P_{\tbox{$\infty$}}(r)$ 
via an auto-convolution of the PRT of $P_{\tbox{E}}(r)$. 
In the extended perturbative regime the major 
features of the saturation profile are captures 
by the PRT. The differences are mainly in the far tails 
where higher order perturbation theory is essential. 
There are also differences in the small scale details,  
where the non-perturbative mixing is important. 
In the non-perturbative regime the saturation 
profile becomes purely non-perturbative, and 
the PRT becomes useless. This is because there 
is no longer separation of energy scales, which is 
the working assumption of the core-tail theory.


We thank Felix Izrailev for suggesting to study the 2DW model. 



\begin{thebibliography}{99}

\vspace*{-1cm}

\bibitem{mario} 
M. Feingold and A. Peres, Phys. Rev. A {\bf 34} 591, (1986).
M. Feingold, D. Leitner, M. Wilkinson, Phys. Rev. Lett. {\bf 66}, 986 (1991). 


\bibitem{wigner} 
E. Wigner, Ann. Math {\bf 62} 548 (1955); {\bf 65} 203 (1957);
Y.V. Fyodorov, O.A. Chubykalo, F.M. Izrailev and G. Casati, 
Phys. Rev. Lett. {\bf 76}, 1603 (1996).


\bibitem{felix} 
G. Casati, B.V. Chirikov, I. Guarneri, F.M. Izrailev, 
Phys. Rev. E {\bf 48}, R1613 (1993); \ Phys. Lett. A {\bf 223}, 430 (1996).  


\bibitem{slfex} 
F. M. Izrailev, T. Kottos, A. Politi and G. P. Tsironis, Phys.~Rev.~E., 
{\bf 55}, 4951 (1997).


\bibitem{smpl} In most of the RMT literature (including the later 
works by Wigner himself), it is assumed that for the purpose 
of `quantum chaos' studies one can consider full (rather than banded) 
matrices, and the first term ${\bf E}_0$ is generally neglected. 
In spite of these enormous simplifications, it turns out that the 
so-called Gaussian invariant ensembles (GOE,GUE) provide a valid 
description of some major spectral properties. 


\bibitem{rsp} 
D. Cohen and T. Kottos, 
Phys. Rev. Lett. 85, 4839 (2000). 


\bibitem{lds} 
D. Cohen and T. Kottos, 
Phys. Rev. E, {\bf 63} 36203, (2001).


\bibitem{ewbrm} Note that the standard BRM model, 
unlike our EBRM model, involves an additional simplification. 
Namely, in case of the former model one further assumes 
that $\mbf{B}$ has a rectangular band profile. 


\bibitem{crs} 
D. Cohen, Phys. Rev. Lett. {\bf 82}, 4951 (1999); 
Ann. Phys. {\bf 283}, 175-231 (2000).
D. Cohen and E.J. Heller, Phys. Rev. Lett. {\bf 84}, 2841 (2000).
D. Cohen, A. Barnett and E.J. Heller, Phys. Rev. E {\bf 63}, 46207 (2001).
D. Cohen, {\em in} Proceedings of the International School
of Physics `Enrico Fermi' Course CXLIII 
``New Directions in Quantum Chaos'', 
Edited by G. Casati, I. Guarneri and U. Smilansky, 
IOS Press, Amsterdam, 2000.  
 
 
\bibitem{wbr} 
D. Cohen, F.M. Izrailev and T. Kottos,  
Phys. Rev. Lett. {\bf 84} 2052 (2000). 


\end{thebibliography}
\end{document}